\documentclass[RNAAS]{aastex62}

\newcommand{\figref}[1]{Fig.\,\ref{#1}}

\newcommand{\cnce}{55\,Cnc\,e}

\newcommand{\Zenv}{$Z_{\rm gas}$\xspace}
\newcommand{\Lenv}{$L_{\rm int}$\xspace}
\newcommand{\rc}{$r_{\rm core}$\xspace}
\newcommand{\rsolid}{$r_{\rm core+mantle}$\xspace}
\newcommand{\menv}{$m_{\rm gas}$\xspace}

\newcommand{\fesima}{{\rm Fe}/{\rm Si}_{\rm mantle}\xspace}
\newcommand{\mgsima}{{\rm Mg}/{\rm Si}_{\rm mantle}\xspace}

\usepackage{soul}
\usepackage{amsmath}
\usepackage{multirow}
\usepackage{xspace}

\submitjournal{RNAAS}

\shorttitle{Update on ``Mass, radius, and composition of the transiting planet \cnce''}
\shortauthors{Crida et al.}
\begin{document}

\title{Mass, radius, and composition of the transiting planet \cnce\,:\\
  using interferometry and correlations\\
--- a quick update}

\correspondingauthor{Aur\'elien Crida}
\email{crida@oca.eu}

\author{Aur\'elien Crida}
\affiliation{Universit\'e C\^ote d'Azur / Observatoire de la C\^ote d'Azur --- Lagrange (UMR\,7293),\\
Boulevard de l'Observatoire, CS 34229, 06300 Nice, \textsc{France}}
\affiliation{Institut Universitaire de France, 103 Boulevard Saint-Michel, 75005 Paris, \textsc{France}}

\author{Roxanne Ligi}
\affiliation{INAF-Osservatorio Astronomico di Brera, Via E. Bianchi 46, I-23807 Merate, Italy}

\author{Caroline Dorn}
\affiliation{University of Zurich, Institut of Computational Sciences,\\
University of Zurich, Winterthurerstrasse 190, CH-8057, Zurich, \textsc{Switzerland}}

\author{Francesco Borsa}
\affiliation{INAF-Osservatorio Astronomico di Brera, Via E. Bianchi 46, I-23807 Merate, Italy}

\author{Yveline Lebreton}
\affiliation{LESIA, Observatoire de Paris, PSL Research University, CNRS, Universit\'e Pierre et Marie Curie, Universit\'e Paris Diderot, 92195 Meudon, \textsc{France}}
\affiliation{Institut de Physique de Rennes, Université de Rennes 1, CNRS UMR 6251, 35042 Rennes, \textsc{France}}




\section{Introduction} 

In a recent paper \citep[][accepted on April 19, hereafter
  C18]{Crida+2018}, we presented a method to derive the mass and
radius of a transiting exoplanet and their intrinsic
correlation. Measuring the stellar radius by interferometry and the
stellar density by analysis of the transit lightcurve allows the
derivation of the stellar mass, independently of any stellar evolution
model. The planetary mass and radius are derived through their ratios
with the stellar ones, given by the transit depth and the amplitude of
the radial velocity signal $K$, whose uncertainties degrade the
correlation. We wrote\,: ``\emph{More precise observations of the
  transit would be very useful in this particular case and would allow
  to increase significantly the gain on the \emph{[planetary]} density
  precision.}''

Three months later, \citet[][hereafter B18]{Bourrier+2018} published
new observations of the system. Additionally, Gaia's DR2 was released
on April 25, 2018. The purpose of this note is solely to implement
these great new data in our pipeline to provide an up-to-date result
of our model.

\vfill

\section{New derived stellar parameters}

Gaia's DR2 catalog \citep{Gaia+2018} provides $d_\star =
12.590\pm0.012$~pc, making the distance used before
($12.34\pm0.11$~pc) inaccurate.

B18 provide from their analysis of the transit lightcurve
$a/R_\star=3.52\pm0.01$ and $P=0.7365474\pm1.3\cdot10^{-6}$~day, from
which we can derive \citep{Seager-MallenOrnelas-2003}
$\rho_\star\approx\frac{4\pi^2}{G}\frac{(a/R_\star)^3}{P^2}=1.079\pm0.005\,\rho_\odot$,
$7.6$ times more precise than what we used before ($1.084\pm
0.038\,\rho_\odot$).

This gives 
$$R_\star = 0.980\pm 0.016\,R_\odot \ \ ;\ \ M_\star = 1.015\pm 0.051\,M_\odot,$$ 
a bigger star than we thought
($0.958\pm0.018\,R_\odot$ and $0.954\pm0.063\,M_\odot$ in C18).
The mass-radius correlation is now $0.995$ versus $0.85$ in C18, due to
the amazing precision on $a/R_\star$, thus on $\rho_\star$.

\newpage

\section{Planetary parameters}

Using these new stellar parameters together with $K=6.02\pm0.24$~m/s
and $R_p/R_\star = 0.0182\pm0.0002$ (B18) instead of $6.30\pm0.21$~m/s
and $0.0192\pm0.0078$ in C18, we find
$$R_p = 1.947\pm 0.038\,R_\oplus\ \ ;\ \ M_p = 8.59\pm 0.43\,M_\oplus\ ,$$
with a correlation $c = 0.54$ and
 $$\rho_p = 1.164\pm 0.062\,\rho_\oplus\ = \ 6421 \pm 342\ {\rm kg.m}^{-3}\ ,$$ 
instead of $2.023\pm0.088\,R_\oplus$ , $8.70\pm0.48\,M_\oplus$ , $c=0.3$ and
$1.06\pm0.13\rho_\oplus\ $.

\begin{figure}
\includegraphics[width=\textwidth]{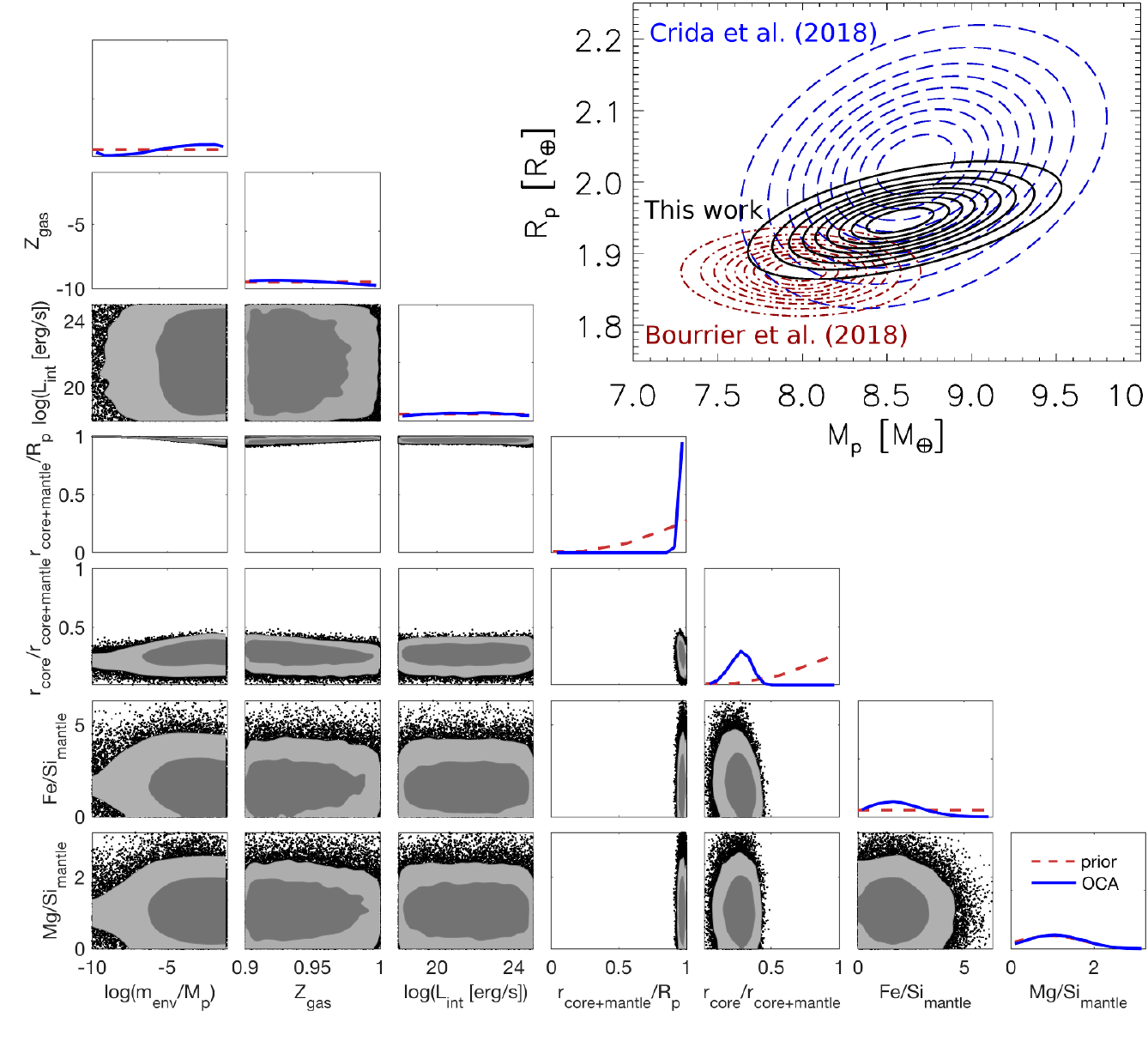}
\caption{Parameters of \cnce. Upper right inset\,: joint PDF of the
  planetary mass and radius.\\ Other plots\,: Sampled two and
  one-dimensional marginal posterior for all interior parameters, same
  as figure~5 of C18.}
\label{figup}
\end{figure}

The joint PDF of $R_p$ and $M_p$ is shown in the inset of \figref{figup}
as black plain line contours. For comparison, the ones from C18 and
B18 are shown as blue long-dashed and red dash-dotted contours
respectively.

\ \\

The other planetary parameters are derived using the model I by
\citet{Dorn-etal-2017a} as in our previous paper in the case
\texttt{OCA}. Their PDFs are shown in \figref{figup}.  Here, we
have used a more restricted prior on the gas metallicity
($0.9<\,$\Zenv$<1$) that excludes hydrogen-dominated gas \emph{a
  priori}. We find\,:
\begin{center}
  \begin{tabular}{rcl}
log$_{10}$({\menv/M$_p$}) & = & $-3.7_{-2.5}^{+1.8}$\vspace{3pt}\\
\Zenv & = & $9.4_{-0.3}^{+0.3}$\vspace{3pt}\\
log$_{10}$(\Lenv) & = & $21.6_{-2.2}^{+2.2}$\vspace{3pt}\\
$r_{\rm gas}$/R$_p$ & = & $0.03_{-0.02}^{+0.02}$\vspace{3pt}\\
\rsolid /R$_p$ & = & $0.97_{-0.02}^{+0.02}$\vspace{3pt}\\
\rc/\rsolid & = & $0.27_{-0.08}^{+0.07}$\vspace{3pt}\\
$\fesima$ & = & $1.79_{-1.03}^{+1.22}$\vspace{3pt}\\
$\mgsima$ & = & $1.1_{-0.6}^{+0.7}$\\
  \end{tabular}
\end{center}

These values, using the most up-to-date observations of the system,
should replace the ones published in our previous paper. As expected,
more precise stellar distance \citep[thanks to Gaia,][]{Gaia+2016},
transit depth and stellar density (thanks to B18) allow us to refine our
estimate of all the planetary parameters. The almost twice better
mass-radius correlation in particular doubles the precision on
$\rho_p$, which reaches $5.3\%$ uncertainty.

\ \\

\section{Discussion and conclusion}

We do not find the same planetary mass and radius as B18, although we
use their transit and radial velocity parameters.\footnote{For
  reference, \citet{Bourrier+2018} give\,: $R_p=1.875\pm
  0.029\,R_\oplus$, $M_p=7.99^{+0.32}_{-0.33}\,M_\oplus$
  (uncorrelated) and $\rho_p = 1.212^{+0.078}_{-0.073}\,\rho_\oplus$.}
This is because they use the stellar radius and mass given by
\citet{vanBraun2011} based on the \emph{Hipparcos} distance and
stellar evolution models, while we use \citet{Ligi-etal-2016}'s
angular diameter and Gaia's parallax for the radius, and the measured
stellar density for the mass. Another difference lies in the models
used for the internal structure (model I from \citet{Dorn-etal-2017a}
here, and II in B18), but this only has a minor impact on the gas
layer thickness.

The method developed in C18 proves very powerful for exoplanets
transiting bright stars\,: observational data provide directly the
mass, radius and density of \cnce, with a precision that allows to
constrain significantly its internal structure. The observations by
Gaia and B18 being at the best possible precision to date, our
numbers can hardly be improved.

\bibliographystyle{aasjournal}
\bibliography{./Cnc.bib}

\end{document}